\documentclass[preprint,showpacs,preprintnumbers,amsmath,amssymb,groupedaddress]{revtex4}
\usepackage{graphicx}
\usepackage{bm}
\usepackage[mathlines]{lineno}
\usepackage{amsmath}
\usepackage{bm}
\usepackage{epsf,epsfig,graphics}
\usepackage{float}
\usepackage{makeidx}
\usepackage{latexsym}
\usepackage{amssymb}
\usepackage{pict2e}
\usepackage{amsfonts}
\usepackage{mathrsfs}
\usepackage{dcolumn}
\usepackage{amsmath}
\usepackage{color}
\usepackage{subfigure}

\begin{document}
\title{Anticipation and Negative Group Delay in a Retina}

\author{Po-Yu Chou$^1$}
\author{Jo-Fan Chien$^{2}$}
\author{Kevin Sean Chen$^{2}$}
\author{\\Yu-Ting Huang$^{2}$}
\author{Chun-Chung Chen$^{2}$}
\author{C. K. Chan$^{1,2}$\footnote{ckchan@gate.sinica.edu.tw}}

\affiliation{
$^1$ Department of Physics, National Central University, Chungli District, Taoyuan 320, Taiwan, R.O.C.\\
$^2$ Institute of Physics, Academia Sinica, Nankang, Taipei 115, Taiwan, R.O.C.}

\pacs{87.19.lj,05.45.Xt,87.19.lt,42.66.Si}
\date{\today}
\begin{abstract}
		
The mechanism of negative group delay (NGD) is used to understand the anticipatory capability of a retina. Experiments with retinas from bull frogs are performed to compare with the predictions of the NGD model. In particulars, whole field stochastic stimulation with various time correlations are used to probe anticipatory responses from the retina. We find that the NGD model can reproduce essential features of experimental observations characterized by the cross correlations between the stimulation and the retinal responses. The prediction horizon of a retina is found to depend on the correlation time of the stimulation as predicted by the NGD model. Experiments with dark and bright Gaussian light pulses further support the NGD mechanism; but only for the dark pulses indicating that the NGD effect of a retina might originate from its OFF response. Our finding suggests that sensory systems capable of using negative feedback for adaptation can give rise to anticipation as a consequence of the delay in the system.
		
\end{abstract}
	
\maketitle
	
%\section{Introduction}
% What is the issue
Anticipation  \cite{stepp2010strong} is a process in which a system generates responses ahead of the actual occurrence of events in the incoming stimulation. For physical systems, a device can produce anticipatory responses when there is negative group delay (NGD) \cite{voss2016signal}. At first sight, anticipation might seem to violate causality but there is a requirement that the signal should be correlated; meaning that the signal can be predicted from its past \cite{hyndman2018forecasting}. NGD devices have been fabricated for fast communication applications where NGD of transmitted signals can improve performance of the system \cite{broomfield2000broadband}. There are plenty of evidence that biological systems possess anticipatory capabilities \cite{stepp2010strong}. It is believed that biological systems make use of anticipation to compensate for the delay in signal processing and propagation in neural systems \cite{stepp2009anticipation}. Even baseball games \cite{mann2013head} can be shown to be related to anticipation. However, it is not clear whether the observed anticipations in biological systems are the result of complex information operations of neural circuits in the brain or just simply an NGD effect.

It is known for a long time that visual systems can produce anticipatory illusion \cite{nijhawan2002neural}. Only recently, Berry et al \cite{Berry1999} and Shwartz et al \cite{Schwartz2008} demonstrated that anticipation can start as early as in the sensor; namely the retina. In the phenomenon of omitted stimulation response (OSR) \cite{Schwartz2008}, a retina was shown to anticipate missing incoming pulse by endogenously generating a response with appropriate timing after a periodic pulse stimulation is abruptly stopped. Chen et al \cite{Chen2017} showed in an extension to OSR that a retina can also distinguish between stochastic pulses generated by a hidden Markov model (HMM) and an Ornstein–Uhlenbeck (OU) process by producing anticipatory responses only for the HMM signal. Since a HMM signal should have also elicited anticipatory responses from a NGD filter, it is possible that the retina is behaving like a NGD devices. One important ingredient for a system to possess anticipatory capability is delayed negative feedback (DNF). Voss \cite{voss2016signal} has shown that DNF can directly lead to NGD. Contrast and gain controls are common for retinas and presumably they are accomplished by mechanism of DNF similar to other control systems \cite{pyragas2001control}. It is highly likely that nature makes use of NGD to endow a retina with anticipatory capability. 

Here, we propose that the NGD mechanism can be used to understand the anticipatory properties of retina and test this idea by both constructing an NGD model and performing experiments with bull frog's retinas in a multi-electrode array (MEA) system. We are interested in comparing the responses from the retina with those from the NGD model when driven by the same stimulation. Similar to Ref\cite{voss2016signal}, we use the time lag cross-correlation to quantify anticipation in the system. Comparison with the prediction of the NGD model shows that some of the operations of a retina can indeed be understood as a NGD device. Furthermore, parameters used in the NGD model is physiologically plausible. 
 
%\section{Material and Methods}
In the original NGD model of Voss\cite{voss2016signal}, the response $y(t)$ of a system driven by input $x(t)$ through a delayed feedback is given by: $\dot y(t) = -\alpha y(t) + k[x(t)- y(t-\zeta)]$ where $\alpha$ and $k$ are the relaxation rate and the gain of the system while $y(t-\zeta)$ is the delayed feedback of $y$ from a time $\zeta$ earlier. This form of NGD model needs storage of $y(t)$ which might not be physiologically feasible. Here we consider the delayed  feedback as coming from another variable $z(t)$ which is a low pass version of $y(t)$ as:
\begin{eqnarray}
\dot y(t) = -\alpha y(t) + k[x(t)- z(t)]\\
\dot z(t) = -\beta z(t) + g y(t)
\end{eqnarray}
\noindent
where $\beta$ and $g$ are defined similarly to $\alpha$ and $k$. With this form, there is no need for the storage. Similar to Ref\cite{voss2016signal}, with $X(\omega)$ and $Y(\omega)$ being the Fourier transforms of $x(t)$ and $y(t)$ respectively, the group delay of the system can be obtained as $\delta(\omega) \equiv -d\Phi(\omega)/d\omega$ with $\Phi(\omega)$ being phase of the response function defined as: $H(\omega) = Y(\omega)/X(\omega) \equiv G(\omega) e^{i\Phi(\omega)}$ at frequency $\omega$. From Eqn(1) and Eqn(2), $\Phi$ can be computed as as:
\begin{eqnarray}
%G(\omega) = \sqrt{\frac{k^2 \left(\beta^2+w^2\right)}{2 \alpha \beta g k+g^2 k^2+\beta^2 %w^2-2 g k w^2+w^4+\alpha^2 \left(\beta^2+w^2\right)}}\\
\Phi(\omega) = -\text{ArcTan}\left[\frac{w \left(\beta^2-g k+w^2\right)}{\beta g k+\alpha \left(\beta^2+w^2\right)}\right]
\end{eqnarray}
\noindent

\begin{figure}[h!]
	\begin{center}
		\includegraphics[width=10cm]{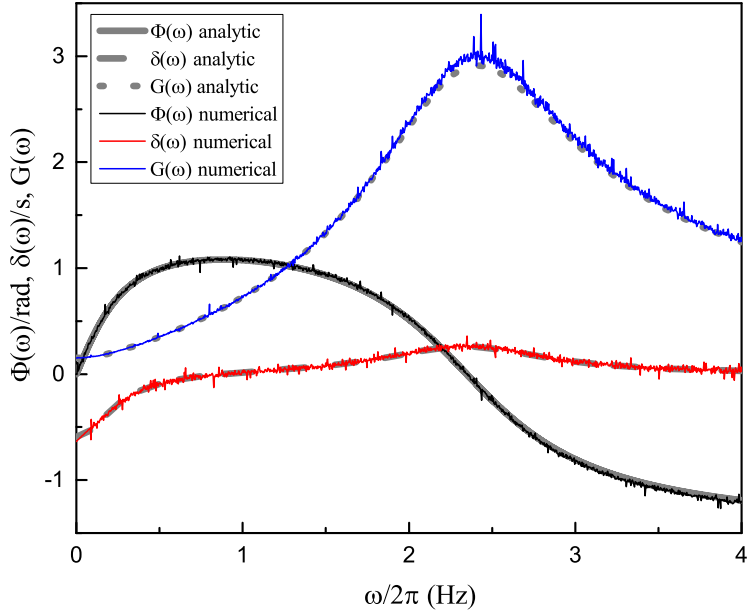}
	\end{center}
\caption{Frequency dependence of gain ($G(\omega)$), phase ($\Phi(\omega)$) and group delay ($\delta(\omega)$) of the response function from the NDG model with $\alpha = 6 s^{-1}$, $\beta = 1.6 s^{-1}$, $k = 22$ and $g= 10$. Both analytic and numerical results are shown. Numerical data points are obtained from numerical simulation described in the text.} 
	\label{f_pd}
\end{figure}

Figure~\ref{f_pd} shows the $\omega$ dependence of $\delta(\omega)$, $G(\omega)$ and $\Phi(\omega)$ with $x(t)$ generated from a time series of random numbers with lowpass filtering similar to \cite{voss2016signal}. The values of the parameters used to produce Figure~\ref{f_pd} is chosen to match observations from the experiments as explained below. From Figure~\ref{f_pd}, a negative $\delta(\omega)$ can be obtained only when $\omega$ is small enough. This is consistent with the idea that only signals with long enough correlation can be predictable. An interesting feature in the figure is that there is a maximum in $G(\omega)$ at about 2.5 Hz. This will show up as oscillations in responses of the system.  Note that Eqn(2) can be solved as: $z(t) = \int_{-\infty}^{t}K(t-t')y(t')dt'$ with the kernel: $K(t-t') = ge^{-\beta(t-t')}$. The case of Voss model can be recovered by setting $K(t) = \delta(t-\xi)$. Eqn(1) and the kernel form of $z(t)$ can be then considered as a generic form for anticipatory dynamics; with $z(t)$ being the convolution of the output $y(t)$ with some kernel $K(t)$ to provide delay. Note that if there is no delay in the system, there will be no NGD effect. We will refer to this model as the NGD model.

Next, we setup experiments to test if this generic form of anticipation dynamics can also be found in a retina. The experiment setup and procedures are similar to Ref\cite{Chen2017}. Retinas used in the experiments were obtained from bullfrogs which were dark adapted for 1 hour before dissection. A small patch of retina tissue was then cut and fixed on a 60-channel multi-electrode array (Qwane Bioscience) with electrodes 10 $\mu$m in diameter spaced at 200 $\mu$m. The retina is perfused with oxygenated Ringer's solution\cite{Ishikane2005} at a rate of 1 ml/min. Each retina was experimented at room tempearture within 6 hours after dissection. We use smoothed Ornstein–Uhlenbeck (OU) time series to generate stimulation. The OU time series $\{s_i\}$ is first generated with: $s_{i+1}  =  (1-\frac{\Delta t}{\tau})s_i + \xi_i \sqrt{D\Delta t}$ where the time step $\Delta t$ is 10 ms, $\xi$ a white noise with unit amplitude, $D = 4 s^{-1}$ the amplitude of the noise and $\tau$ the relaxation time of the system. Next, stimulation $\{x_i\}$ with various time correlations, are generated from $\{s_i\}$ by using low-pass filters with different cutoff frequencies ($f_c$). The time series $\{x_i\}$ are then used to control the light intensity $I(t) $ of an LED (peak of wavelength = 560 nm) to stimulate the whole retina. The maximum  and minimum light intensities used are  $18$ and $2 mW/m^2 $ respectively with average intensity of $10 mW/m^2$. Responses from the retina ($\{r_i\}$) are recorded by the MEA as a function of different $f_c$ and $\tau$. 

In a typical successful experimental recording, about 70\% of the MEA  electrodes are generating responses. For the responding electrodes, they usually give different responses. Presumably, this diversity is due to the existence of different pathways in a retina \cite{famiglietti1977neuronal}. The spikes obtained are spikes sorted to remove redundant detection. In Ref\cite{voss2016signal}, cross-correlation function (XCF) between stimulation and response $\langle x(t)y(t+\delta t)\rangle_t$ at different time delay $\delta t$  are used to characterize the anticipatory property of the system. If there is a peak in $\langle x(t)y(t+\delta t)\rangle_t$ at location $\delta t = \delta t_p$, a positive $\delta t_p$ signifies that $y(t)$ is anticipatory of $x(t)$. These XCF are similar to the spike triggered average (STA)\cite{paninski2006spike} which characterizes the averaged wave form before the generation of a spike at $\delta t = 0$. In the experiments, the forms of the STA obtained from different channels can roughly be divided into two groups; namely the predictive (P-channel) and non-predictive (NP-channel) as explained below. Since behaviors of every retina can be quite different in details and cannot be averaged, we are reporting the behaviors of a single retina below. But the reported behaviors are representative of more than 10 retinas from 10 different animals.

%\begin{figure}[h!]
%	\begin{center}
%		\includegraphics[width=8cm]{NGDfig/Fig_MI_Exp}
%	\end{center}
%	\caption{Retinal responses to whole field OU stimulation. a) Computed TLMI as a function of time lag $\delta t$ under OU stimualtion with various upper cut-off frequency $f_c$ for a predictive channel. There are two groups of peaks. The peak on the right are predictive peaks as the peak positions shifted towards the future (positive $\delta t$) when $f_c$ is lowered. b) The corresponding spike triggered average $h(t)$ of experiments in a).}
%	\label{f_MI_Exp}
%\end{figure}

\begin{figure}[h!]
	\begin{center}
		\includegraphics[width=14cm]{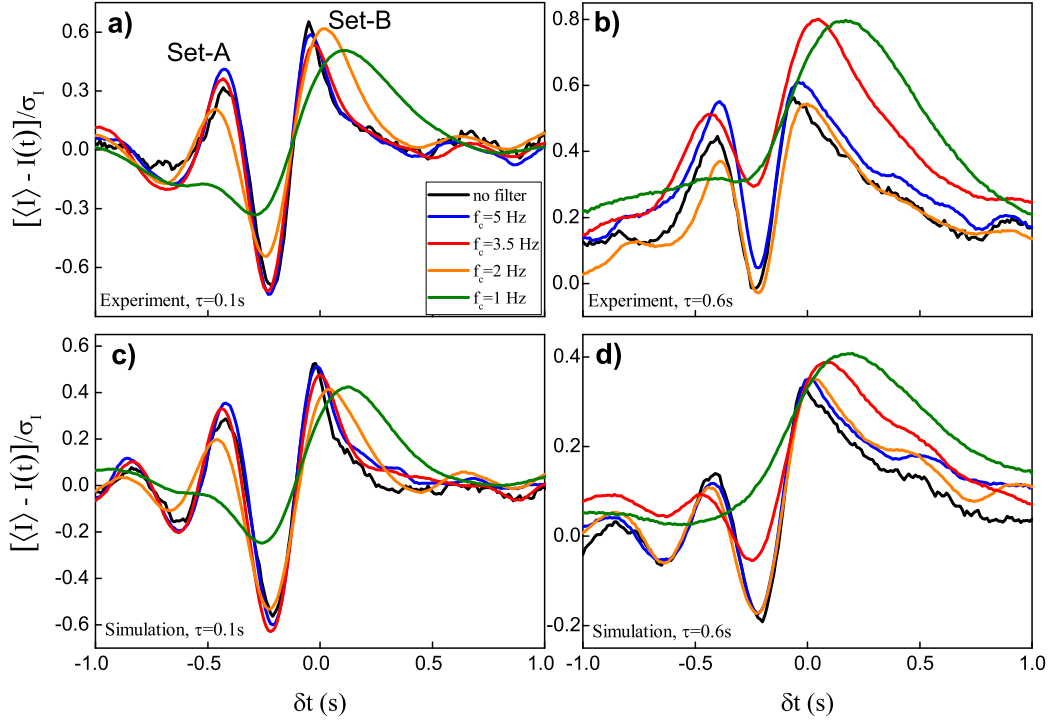}
	\end{center}
	\caption{Comparison of spike triggered average (STA) obtained from experiments and simulation with various $f_c$ and $\tau$: a) and b) are the inverted STA from experiment with $\tau = 0.1$ and $0.6 s$ respectively. Note that the oscillatory nature of the iSTA gives rise to two sets of peaks in the iSTA which are marked as Set-A and Set-B. c) and d) are the corresponding STA generated from the NDG model with the same parameters of Figure 1 with the same stimulation used in a) and b). The data are obtained from experiments in which stimulation lasted for 300 s. }
	\label{f_iSTA}
\end{figure}

Figure~\ref{f_iSTA}a shows the inverted STA (iSTA = -STA) obtained from a typical P-channel for two different $\tau$ with various $f_c$. We need to use iSTA here because the NGD model will produce results similar only to iSTA of a P-channel as shown below. A remarkable feature of Figure~\ref{f_iSTA}a is that there are two sets of peaks in the iSTA as marked Set-A and Set-B in the figure. Note that the peak positions ($\delta t_p$) of the iSTA in Set-B are shifted towards more positive $\delta t$ when the $f_c$ are reduced while those of Set-A are shifted in the opposite direction. Anticipatory behaviors can be inferred from some of the peaks in the Set-B peaks which are located at $\delta t_p >0$ when their corresponding $f_c$ are small enough. This is the reason for labeling this kind of responses as predictive. No such predictive behaviour can be found from the NP-channel (Figure 1 in Supplementary Materials (SM)). The effect of correlation on anticipation found here is similar to those of Ref\cite{Chen2017} by the method of time lag mutual information (TLMI); namely prediction is possible only for stimulation with long enough correlation time or small enough $f_c$. In fact, similar conclusion can also be reached with the method of TLMI (Figure 2 in SM).
 
To test if the NGD model can capture essential features of the experiments, we have performed simulation of the NGD model with the same stimulation from the experiment by adjusting the parameters $(\alpha, \beta, k, g)$ to best match with the shapes of experimental iSTA. Figure~\ref{f_iSTA}c) and d) shows the corresponding STA from the NGD model. These STA are generated by first passing  $y(t)$ generated from the NGD model through an activation function and then used the activated output ($\lambda(t)$) to generate a Poisson spikes train with a firing rate proportional to the $\lambda(t)$. These model generated STAs are very similar to those of the iSTA from experiments. For a retina, an OFF pathway \cite{famiglietti1977neuronal} generates positive response when there is an decrease in the input light intensity and similarly for an ON-pathway. In this sense, the NGD model is an ON pathway. 
Presumably, the spikes recorded from the P-channels must be predominately from the OFF-path way and this is the reason we need to use iSTA for a P-channel.

\begin{figure}[h!]
 	\begin{center}
 		\includegraphics[width=10cm]{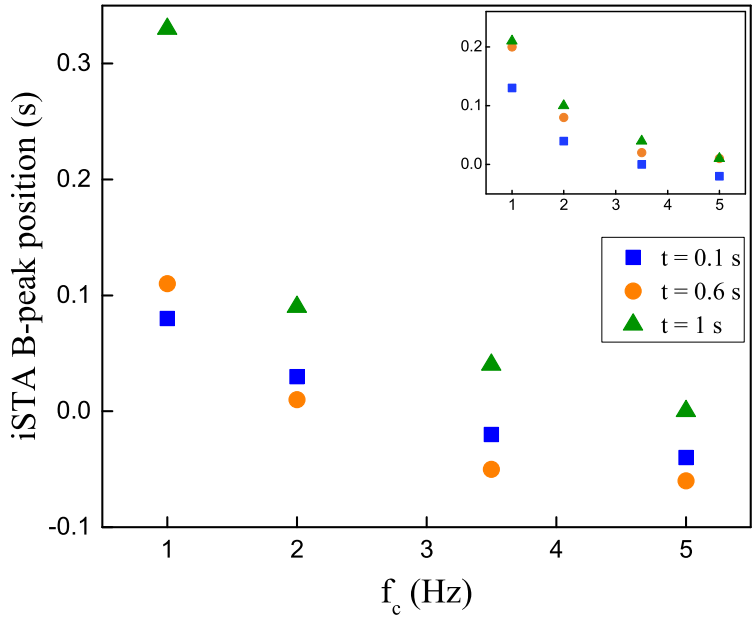}
 	\end{center}
 	\caption{Peak positions ($\delta t_p$) of peaks in Set-B of the iSTA or STA as a function of cut-off frequencies for three different values of $\tau$ from both experiments and simulation (inset). Note that a positive value of $\delta t_p$ indicates anticipatory response from the retina or the NGD model.}
 	\label{fig_ts}
 \end{figure}

Note that Set-A and Set-B peaks are also reproduced by the simulation. Similar to the experiments, $\delta t_p$ of peaks in Set-B from simulation are also shifted towards more positive $\delta t$ when $f_c$ are lowered. The $f_c$ and $\tau$ dependence of $t_p$ of Set-B peaks for both experiments and simulations are shown in Figure~\ref{fig_ts}. In the figure, $\delta t_p$ from both experiments and simulation show similar dependence on $f_c$ but not for the $\tau$ dependence. Presumably, the retina might have dynamics which are not captured by the NGD model. Furthermore, both STAs from experiments and simulation show oscillations with a characteristic frequency consistent with the peak position of $G(\omega)$ in Fig~ \ref{f_pd}. All the simulations results reported here are generated with only one single set of parameters; indicating that the simulation can mimic the experiment at least for the $f_c$ dependence.

\begin{figure}[h!]
	\begin{center}
		\includegraphics[width=12cm]{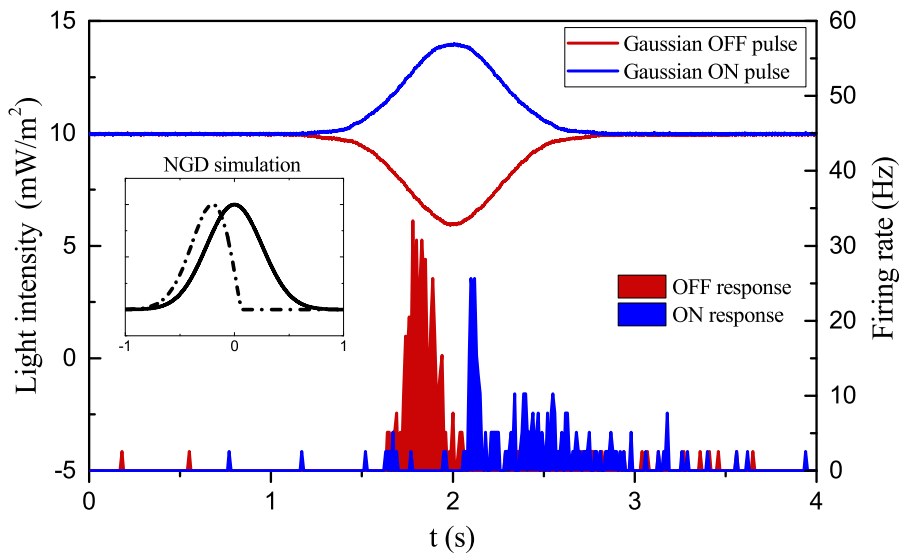}
	\end{center}
	\caption{Responses of a retina and the NDG model to Gaussian pulses (see main text for details). The red and blue traces are for the bright (ON) and dark (OFF) pulses stimulation respectively in the experiments. The corresponding responses from the retina measured as firing rate are also shown with respective colors. The data is obtained by averaging the responses from the same retina over 40 runs. The P-channel used here is the same channel used in Figure 2.  Inset shows the anticipatory response (broken line) of the NGD model with same parameters used to generate Figure 2c and 2d for an incoming Gaussian pulse (black line).}
	\label{fpulse}
\end{figure}

One of the hall marks of a NGD system is the generation of time advanced pulse \cite{kitano2003negative}. To further test the NGD properties of the retina, Gaussian pulses are also used to test if the pulses can be advanced by the retina. Figure~\ref{fpulse} shows the responses of the retina for both an ON and an OFF pulses in the form: $S(t) = I_0  \pm  I_p e^{-\frac{1}{2}(t/ 2 \sqrt{2ln(2)}\tau_w)^2}$  with $\tau_w$ being the  half maximum width of the pulse. The reason of using both ON and OFF pulses is that the retina can response to both increasing and decreasing light stimulation in its ON and OFF pathway \cite{famiglietti1977neuronal}. Figure~\ref{fpulse} shows that the ON pulse is delayed in the response of retina while that from the OFF pulse is advanced. It seems that the retina is an NGD device only for the OFF pulses. The pulse response from an NP-channel is very different (see SM for details). The simulation of a pulse input has also been performed with the NGD model as shown in the inset with the same set of parameters in Figure~\ref{f_pd}; showing agreement with experiments. Figure~\ref{fig_ts} shows that the P-channels are capable of producing both ON and OFF spikes. The same is also true for the NP-channel (Figure 3 of SM). From both of these P-channel and NP-channel recordings, it can be seen that it is the OFF responses which are predictive of the incoming pulses. However, for faster stimulation, as in 
Figure~\ref{f_iSTA}a and \ref{f_iSTA}b for a P-channel, it is not clear why most of the spikes are from the OFF-pathway and therefore rendering the P-channel predictive. 

%As mentioned above, not all the channels from the MEA are predictive. The ON and OFF pulses results suggest that the P-channels can produce both ON and OFF response. A detailed check with ON and OFF tests (see SFig.1 in SI) shows that indeed the ON channels are non-predictive and vice versa for the OFF channels. A remarkable feature of the spike trains measured from the ON and OFF channels is that there are many common spikes between these two spike trains as shown in inset of Figure 5. Therefore, if $P$ and $N$ are the sets of spike from the P and NP-channels channels respectively, one has: $P = P'\cup P"$ and $N = N' \cup N"$ where $N'=P'$ are the shared spikes and $P"$ ($N"$) are unique spikes in $P$ ($N$). Spike triggered average \cite{Chichilnisky2001} analysis of $P"$ and $P'$ shows that $P'$ are from ON pathway vice versa for $P"$. Once the spikes are separated into ON and OFF pathways, we can perform TLMI analysis again with only the the ON or OFF spikes as shown in Figure 5 which shows that the first peak in Figure 1 comes from the ON spikes while the OFF spikes contribute to the second peak; suggesting that it is the OFF spike which can generate anticipatory spikes.

From the discussions above, it is clear that the extended Voss NDG model can capture essential features of the anticipatory behaviors of a retina. The essence of the NDG mechanism is a delayed negative feedback and it is well known that there is delayed feedback from the horizontal cell to the photo-detector cells (cone cells) in a retina. The rate constants $\alpha$ and $\beta$ in the model can be related to the relaxation time constants of the cone ($\tau_y$) and horizontal ($\tau_h$) cells respectively. Drinneberg et al \cite{drinnenberg2018diverse} studied the effect of this feedback with a retina from a rat also under whole field stimulation. The rate constants $\tau_y^{-1} = 19.8 s^{-1}$ and $\tau_h^{-1} = 2,7 s^{-1}$ used in Ref\cite{drinnenberg2018diverse}) are of the same order of magnitude of $\alpha = 6 s^{-1}$ and $\beta = 1.6 s^{-1}$ used in the NGD model for a frog; suggesting that the NGD model is biologically feasible. In fact, the $z(t)$ in the NGD model is just the convoluted output of the cone cell produced by the horizontal cell in Ref\cite{drinnenberg2018diverse}.

The anticipatory model described by Eqn(1) and Eqn(2) can also be considered as an adaptation model for a sensor to generate response ($y(t)$) to inform downstream system of the changes in the environment ($x(t)$). The term $[x(t)-z(t)]$ can be understood as the "error" between the incoming signal $x(t)$ and an adaptive internal reference $z(t)$. The action of the model is simply to adjust its output $y(t)$ to adapt to the environment by minimizing this "error" signal. Interestingly, it is precisely the delay in the system which gives rise to the anticipatory response in $y(t)$. Consequently, the prediction horizon of $y(t)$ will be lengthened when there is longer delay. In the concept of perceptual control \cite{powers1973behavior}, the observed anticipatory response of a retina is simply the result of the control of perception ($y(t)$) of the external world ($x(t)$) by the retinal circuit through a delayed negative feedback mechanism. Since biological sensors should have the ability to adapt to changes in their environment through negative feedback, presumably anticipatory capabilities should be found in a wide varieties of biological sensing systems. 

This work has been supported by the MOST of ROC under the grant number 108-2112-M-001 -029 -MY3.

%\bibliography{SBRef.bib}
\bibliographystyle{prsty}
\bibliography{retina_ref}

\newpage
	
	% Figure 1

\end{document}